# Surface Electromyography-controlled Pedestrian Collision Avoidance: A Driving Simulator Study

Edric John Cruz Nacpil[a], Zheng Wang[a]*, Zhanhong Yan[a], Tsutomu Kaizuka[a], and Kimihiko Nakano[a]

[a]*Institute of Industrial Science, The University of Tokyo, Tokyo, Japan;*

*z-wang@iis.u-tokyo.ac.jp

Funding: This work was supported by a Grant-in-Aid for Early-Career Scientists (no. 19K20318) from the Japan Society for the Promotion of Science.

# Surface Electromyography-controlled Pedestrian Collision Avoidance: A Driving Simulator Study

Drivers with disabilities such as hemiplegia or unilateral upper limb amputation restricting steering wheel operation to one arm could encounter the challenge of stabilizing vehicles during pedestrian collision avoidance. An sEMG-controlled steering assistance system was developed for these drivers to enable rapid steering wheel rotation with only one healthy arm. Test drivers were recruited to use the Myo armband as a sEMG-based interface to perform pedestrian collision avoidance in a driving simulator. It was hypothesized that the sEMG-based interface would be comparable or superior in vehicle stability to manual takeover from automated driving and conventional steering wheel operation. The Myo armband interface was significantly superior to manual takeover from automated driving and comparable to manual steering wheel operation. The results of the driving simulator trials confirm the feasibility of the sEMG-controlled system as a safe alternative that could benefit drivers with the aforesaid disabilities.

Keywords: automated driving; pedestrian collision avoidance; collision avoidance systems; steering assistance; sEMG

## 1. Introduction

Drivers could have less than half a second to avoid pedestrians who unexpectedly traverse roads. If the time-to-collision is 0.4 s, steering has been observed through field testing to be more effective than braking at successful pedestrian collision avoidance (Adams, 1994; Barrett et al., 1968). Although steering systems have generally been developed with collision avoidance to assist drivers without disabilities, there is a paucity in the literature on pedestrian collision avoidance to assist drivers with upper limb disabilities (Dang et al., 2012). The current study contrastingly addresses drivers with unilateral upper limb amputation or hemiplegia by proposing a sEMG-controlled collision avoidance system (CAS) that utilizes the Myo armband as a sEMG-based interface device. The CAS was designed to safely steer a vehicle based on input from the driver. A driving simulator experiment involving human test drivers validated the safety of the CAS with respect to vehicle stability, i.e. vehicle slip angle. It was hypothesized that the vehicle stability of the Myo armband interface would be comparable or superior to the same performance parameters for conventional steering wheel operation and manual takeover from fully automated driving.

The remainder of this paper is structured as follows: Section 2 considers research related to the current study, whereas Section 3 details the methodology for the proposed CAS and driving simulator experiments. Section 4 presents and discusses the experimental results, followed by concluding statements in Section 5.

## 2. Related Work

Various CASs employing steering as a road obstacle avoidance strategy can be classified as: driver-initiated evasion assistance, corrective evasion assistance, and automatic evasion assistance (Dang et al., 2012). Without requiring any steering input from the driver, automatic evasion assistance for fully automated driving relies on advanced sensors to obtain detailed information about the ego vehicle environment to execute avoidance maneuvers, e.g. the frontal obstacle system developed by Hayashi et al. (Hayashi et al., 2017). On the other hand, corrective evasion assistance systems only activate, if a potential collision is detected through vehicle-mounted sensors. However, road obstacle sensors, such as lidar devices, can be costly and environmental conditions such as rainy weather can reduce the ability of these sensors to accurately and comprehensively acquire information (Kukkala et al., 2018).

On the other hand, as more feasible and cost-effective alternative, driver-initiated evasion assistance allows the driver to choose the time at which to begin an automated steering maneuver, as oppose to relying on vehicle sensors. Although manual takeover of the vehicle has been considered as a possible method for dealing with unexpected road scenarios, driving simulator testing has demonstrated the possibility of steering by elderly drivers during manual takeover that could lead to accidents (Manawadu et al., 2018; Umeno et al., 2018). A further challenge is the presence of stroke-induced hemiplegia among elderly drivers, or unilateral upper limb amputation among drivers of different ages, that restricts collision avoidance to one-handed steering with a healthy arm (Hitosugi et al., 2011; Verrall and Kulkarni, 1995). The current research addresses drivers with these disabilities by developing a sEMG-based interface to provide driver-initiated evasion assistance.

A review of the literature yields only a handful of studies over the past two decades concerning the application of sEMG as an interface for controlling vehicle functions. Some studies have developed prototype sEMG equipment for actual automobiles and driving simulators (de Freitas et al., 2019; Kwak et al., 2008; Nacpil et al., 2018). For example, some studies proposed a sEMG-controlled steering assistance interface for performing low-speed, non-emergency turning maneuvers (E. J. C. Nacpil et al., 2019a, 2019b). A more recent study concerns the development of a prototype sEMG-based driving assistance interface for bilateral transhumeral amputees (E. J. Nacpil et al., 2019), whereas another recent study utilized the Myo armband, a mass-produced sEMG-based interface for low-speed routine maneuvers (Nacpil and Nakano, 2020). On the other hand, the current study considers the Myo armband for emergency steering during pedestrian collision avoidance. Hence, the driving simulator experiment detailed in the next section is a significant contribution to existing literature.

## 3. Materials and Methodology

This section details the development of the proposed sEMG-based collision avoidance system. Section 3.1 covers the design of the system for a driving simulator, whereas Section 3.2 describes experimental methods to assess the vehicle stability of the system.

## 3.1 Steering Assistance Interface for Driving Simulator

In order to control the steering wheel angle (SWA), $\delta_H$, of the driving simulator steering wheel, the control scheme in Figure 1 was implemented. Extending the wrist of the right arm generates sEMG signals that were detected by the Myo armband. Based on this detection, a command was sent by the Myo armband to a laptop that translated the command into an ethernet signal that was sent to the driving simulator host computer. Then the host computer sent a command to an electric control unit that provided a voltage signal to a DC motor. Finally, the steering column was rotated by the motor to adjust the SWA.

The driving simulator host computer executed a steering wheel control algorithm based on the work of Wang et al. (Wang et al., 2019). This algorithm implemented a proportional-integral, proportional-derivative controller as expressed by

$$T_h = a_1 e_{y(near)} + a_2 \int e_{y(near)} \, dt + a_3 e_{\theta(far)} + a_4 \dot{e}_{\theta(far)} \qquad (1)$$

The controller incorporates far and near points of the road ahead of the simulated vehicle. The near point centers the vehicle in the lane, whereas the far point enables the controller to adjust to upcoming road curvature. The lateral error, $e_y$, is defined as the distance between the vehicle and the preprogramed target trajectory, with respect to the near point. On the other hand, the yaw error, $e_\theta$, is the angle between the longitudinal path of the vehicle and the preprogrammed target trajectory at the far point. The constant values of $a_1$ through $a_4$ are decided as 0.19, 0.019, 3.8, and 0.19, respectively, based on trial-and-error driving simulations to confirm the ability of the simulated vehicle to avoid a pedestrian. The control method is an instance of SAE Level 3 automation by which the vehicle controls steering, acceleration, and braking, while the driver monitors the vehicle surroundings in order to intervene by using the Myo armband, if necessary (SAE international, 2016). The torque applied to the steering wheel, $T_h$, was limited to 5 N·m so that test drivers could manually correct the SWA at any time (Wang et al., 2017).

**3.2   Methodology**

*3.2.1   Driving scenarios*

Driving scenarios were designed to replicate pedestrian collision avoidance on residential roads in Japan (Barrett et al., 1968). Since allowing test subjects to brake during the driving scenarios could interfere with the measurement of steering trajectories and interface performance variables, programming the simulated vehicle to run on cruise control at 30 km/h, and executing scenarios where steering was more effective than braking, was suited to validating the Myo armband as a steering interface. The time-to-collision of the simulated vehicle with the pedestrian was set to 0.3 s, which is below the empirically observed time-to-collision of 0.4 s for scenarios where steering was more effective than braking due to the brief time frame (Barrett et al., 1968).

   With the test drivers navigating the vehicle on the left lane, as shown in Figure 2, each driving scenario was programmed to have a pedestrian run perpendicularly across the road from behind a parked vehicle on the left-hand side at 8.34 m/s, after the driver was allowed at least 5 min to adjust to the driving task (Chrysler et al., 2015; Frishberg, 1983). The pedestrian came to a full stop, when reaching the center of the left lane. In accordance with pre-experimental training, upon seeing a pedestrian running across the road, the drivers steered the vehicle to the right without pressing the brake.

*3.2.2   Test subject recruitment*

With permission from the Ethics Committee of the Interfaculty Initiative in Information Studies, Graduate School of Interdisciplinary Information Studies, The University of Tokyo (No. 14 in 2017), 12 healthy male test subjects with an average age of 23.3, an average of 3.1 years of actual driving experience were recruited as test drivers. All subjects gave informed consent prior to participating in the experiment. Only test subjects ages 25 and under were recruited, since drivers at or below 25 years of age are more likely to be involved in pedestrian collision accidents (Chrysler et al., 2015). Hence, the recruitment of drivers within this age bracket would more accurately approximate actual pedestrian collision avoidance scenarios.

*3.2.3   Experimental procedure*

As shown in Table I, all test subjects performed seven experimental conditions. Each of the conditions lasted five to six minutes to allow the drivers to become use to the driving task. Within-subject partial counterbalancing of conditions with Latin squares was performed to address learning effects resulting from the order in which the conditions were performed. In addition to the six conditions involving pedestrian avoidance, another condition without collision avoidance was performed as the forth condition across all test subjects, in order to reduce the influence on the test results from the expectation among the test subjects that every driving scenario included collision avoidance.

Training of all test subjects included, as the first scenario, conventional steering wheel operation without any collision avoidance. Subsequent training scenarios all involved pedestrian collision avoidance at the same crosswalk, while the driver used one of three steering interfaces in the following order: steering wheel, Myo armband, manual takeover. In the case of manual takeover, drivers began the training scenario with acceleration and steering on autopilot. When the pedestrian began running across the road, the steering was switched to manual mode so that the driver could manually resume control of the steering wheel. The training scenario for the Myo armband, as well as the experimental scenarios for manual takeover and the Myo armband also began in autopilot mode.

Since there were three interfaces, there were three training scenarios with collision avoidance, and one scenario without collision avoidance, for a total of four training scenarios. For the sake of efficiency, all training scenarios lasted up to 2.5 min.

*3.2.4 Data analysis*

As a basis for comparing the steering interfaces with respect to vehicle stability, the vehicle slip angle, $\beta$, as graphically defined in Figure 3, was measured at 120 Hz (Hac and Simpson, 2000).

In order to eliminate the possibility that recorded vehicle slip angles at or close to zero could skew the average vehicle slip angle towards zero, especially in instances where the vehicle travels along a longitudinal trajectory before steering to avoid the simulated pedestrian, data points less than 0.1º and greater than -0.1º, were excluded from analysis. Since the vehicle slip angle could be either positive or negative, yet the control design of the Myo armband interface was to reduce the absolute value of the vehicle slip angle, the absolute value of the measured vehicle slip angle was calculated. If the Myo armband was, at minimum, comparable to the other interfaces with respect to the absolute average vehicle slip angle across all test subjects, then the Myo armband would be validated.

Before comparing the steering interfaces, it was determined whether or not the presence of a crosswalk affects the vehicle slip angle. For each of the two driving scenarios in Figure 2, a single average absolute vehicle slip angle was calculated across all three interfaces for each of the 12 test subjects for a total of 12 averages per scenario. Since the Shapiro–Wilk test in MATLAB indicated that the average absolute vehicle slip angles from each test subject were normally distributed only for the scenario with the pedestrian at the crosswalk, where $p < 0.05$ (Shapiro and Wilk, 1965), the nonparametric Wilcoxon signed rank test was used in MATLAB to determine whether the driving scenarios were significantly different from each other (Whitley and Ball, 2002).

Since the Wilcoxon signed rank test indicated with $p > 0.05$ that the presence of a crosswalk did not significantly affect the magnitude of the vehicle slip angle, the interfaces were compared with respect to the absolute average vehicle slip angles across the driving scenarios. Each of the 12 test subjects operated all the steering interfaces, thus resulting in 12 absolute average vehicle slip angles per interface. The Shapiro-Wilk test only confirmed the normal distribution of these slip angles for the steering wheel, and thus the Wilcoxon signed rank test was used to compare the interfaces, where $p < 0.05$ to reject the null hypothesis that there was no statistical difference between two given interfaces. In order to graphically compare the interfaces, three box plots were generated in Excel to represent the average absolute vehicle slip angle distributions for each interface.

As a means of explaining the outcome of the interface comparison with respect to vehicle motion, average lateral acceleration over time, as graphically defined in Figure 3, was plotted. The pedestrian begins to cross the road in front of the car at $t = 0$, whereas the final time is the moment before the car passes the location of the pedestrian along the road.

Further explanation of the results is provided by comparing the interfaces with respect to the average minimum distance to the pedestrian, average maximum steering wheel angle, and average response time of the steering wheel angle, i.e. average time to turn the steering wheel 5º to the right. The Wilcoxon signed rank test was used to determine the significance of differences between averages of for these performance parameters that were not normally distributed, whereas averages of normal distributions were compared with the *t*-Test (de Winter, 2013). The *F*-test for equal variances between distributions was used to select the *t*-Test for unequal variances or the *t*-Test for equal variances (Snedecor, 1934; Welch, 1947).

## 4 Results

A comparison of the three tested steering interfaces with regard to the average absolute vehicle slip angle during pedestrian collision avoidance is shown in Figure 4. Regardless of whether or not the pedestrian was using a crosswalk, the Wilcoxon signed rank test indicated that the Myo armband was comparable to manual steering wheel operation and significantly more stable than manual takeover from automated driving. Therefore, since it was hypothesized that the Myo armband would be comparable or superior to the other interfaces, the Myo armband was validated with respect to vehicle stability. The following subsections discuss several vehicle dynamics that contributed to this outcome.

### 4.1 Vehicle Trajectory and Lateral Acceleration

Lateral acceleration on the simulated vehicle is one potential contributing variable to the vehicle slip angle as indicated in Figure 2. One method for explaining the change in lateral acceleration on the simulated vehicle during pedestrian collision avoidance is to observe changes in the collision avoidance trajectories, as shown in Figure 5. Noting that the position of the vehicle was measured relative to the absolute *x-y* origin in the driving simulation scenario map, the vehicle runs parallel to the *x*-axis at the beginning of each driving scenario. In order to avoid the pedestrian, drivers were instructed to steer the vehicle rightward from the left lane, thereby causing the vehicle to transition into the right lane. In the case of conventional steering wheel rotation and manual takeover from automated driving, this lane change was executed, on average, at a farther distance from the pedestrian, than the distance observed in the case of the Myo armband. This indicates a faster response time for conventional steering and manual takeover with regard to steering initiation. However, the less gradual lane change with the steering wheel or manual takeover can be attributed to direct steering wheel rotation by human test drivers.

Steering was more abrupt when the pedestrian was at the crosswalk because the presence of the crosswalk could have caused drivers to steer more rapidly in anticipation of the crossing pedestrian. Since lateral acceleration on the vehicle towards the inside of the turn, i.e. centripetal acceleration, increases as steering becomes more abrupt, Figure 6 indicates manual takeover and conventional steering wheel operation produced higher peak lateral acceleration than the Myo armband.

Increased lateral acceleration can contribute to loss in tire traction, and consequently, vehicle stability during pedestrian collision avoidance (Hac and Simpson, 2000). Negative average lateral acceleration occurs when the vehicle steers away from the pedestrian, i.e. to the right of the driver. When the vehicle steers in the opposite direction to enter the lane for opposing traffic, the lateral acceleration becomes positive. The period during collision avoidance between 1 s and 3 s contains dynamic changes in lateral acceleration that, in turn, changed vehicle stability among the steering interfaces, as shown in Figure 4.

## 4.2 Steering Wheel Angle

The vehicle trajectories in Figure 5 indicate that the Myo armband was associated with the lowest minimum distance between the vehicle and the pedestrian. As listed in Table II, the Myo armband had average minimum distances to the pedestrian of 1.62 SD 0.52 m and 1.11 SD 0.39 m, when the pedestrian was at the crosswalk and when the pedestrian was not at the crosswalk, respectively. These shorter distances are a consequence of the more gradual vehicle trajectories of the Myo armband. Two possible contributors to these trajectories are the longer average response time and maximum value of the steering wheel angle, since the rate and angle at which the steering wheel is turned affects the trajectory of the vehicle. This reasoning coincides with the lower average maximum steering wheel angle of the Myo armband, relative to the other interfaces, as shown in Table 2. Furthermore, in the scenario with the pedestrian at the crosswalk, the Myo armband had an expectedly longer average response time than the other interfaces. However, the steering wheel had the longest response time when the pedestrian was not at the crosswalk. Since the interfaces only differ by hundredths of a second for each driving scenario, it is possible that the rate of change of the steering wheel angle had a negligible effect on the trajectory of the vehicle.

   On the other hand, across all the driving scenarios, Table 3 indicates that the Myo armband was significantly lower than the other interfaces with respect to the average maximum steering wheel angle. Therefore, the results suggest that the magnitude of the steering wheel angle contributed more than response time to vehicle trajectory. Moreover, since vehicle trajectory corresponds to vehicle slip angle, the lower magnitude of the steering wheel angle for the Myo armband could have contributed to increased vehicle stability.

## 5. Conclusions

The safety of the sEMG-based interface, with respect to vehicle stability, was confirmed through its comparison with manual takeover from automated driving and with manual steering wheel operation. Driving simulator testing demonstrates that use of the interfaces tended to result in more gradual collision avoidance trajectories and lower steering wheel angles that were observed to translate into higher vehicle stability.

Although the minimum distance to the pedestrian of the sEMG-based interface was lower than the steering wheel-based interfaces, such a tradeoff was acceptable, since the sEMG-based interface had higher vehicle stability, and was therefore safer than the other tested interfaces. Although healthy test subjects were recruited for the current study, persons with disability could participate in future studies involving a driving simulator or an actual automobile. Future work could also include other dynamic steering scenarios involving vehicle slip, such as avoidance of environmental vehicles, or collision avoidance involving the use of braking. Despite its limitations, the current study supports the implementation of the sEMG-based interface to assist some drivers with disabilities during pedestrian collision avoidance.

Table 1. Conditions for testing steering interfaces during pedestrian collision avoidance.

| Condition Number | Interface | Pedestrian used crosswalk? |
|---|---|---|
| 1 | Myo Armband | Yes |
| 2 | Myo Armband | No |
| 3 | Steering Wheel | Yes |
| 4 | Steering Wheel | No pedestrian |
| 5 | Steering Wheel | No |
| 6 | Manual Takeover | Yes |
| 7 | Manual Takeover | No |

Table 2. Comparison of interfaces with respect to multiple performance parameters averaged over 12 test drivers.

| Performance Parameter | Interface | Pedestrian at Crosswalk | Pedestrian Not at Crosswalk |
|---|---|---|---|
| Average Minimum Distance to Pedestrian (m) | Myo Armband | 1.62 SD 0.52 | 1.11 SD 0.39 |
| | Steering Wheel | 2.73 SD 0.54 | 2.68 SD 0.95 |
| | Manual Takeover | 2.81 SD 0.55 | 3.06 SD 0.45 |
| Average Maximum Steering Wheel Angle (º) | Myo Armband | 39.02 SD 31.37 | 30.02 SD 18.49 |
| | Steering Wheel | 85.65 SD 57.57 | 83.06 SD 53.33 |
| | Manual Takeover | 92.88 SD 43.07 | 101.4 SD 43.4 |
| Average Response Time of Steering Wheel Angle (s) | Myo Armband | 3.74 SD 0.06 | 3.89 SD 0.12 |
| | Steering Wheel | 3.69 SD 0.1 | 3.96 SD 0.12 |
| | Manual Takeover | 3.71 SD 0.08 | 3.90 SD 0.08 |

Table 3. Statistical significance of steering interface performance parameters.

| Driving Scenario | Performance Parameter | Interface | Myo Armband | Steering Wheel | Manual Takeover |
|---|---|---|---|---|---|
| Pedestrian at Crosswalk | Minimum Average Distance to Pedestrian (m) | Myo Armband | – | 0.000*,### | 0.000*,## |
| | | Steering Wheel | 0.000*,### | – | 0.763## |
| | | Manual Takeover | 0.000*,## | 0.763### | – |
| | Maximum Average Steering Wheel Angle (º) | Myo Armband | – | 0.044### | 0.001### |
| | | Steering Wheel | 0.044### | – | 0.303### |
| | | Manual Takeover | 0.001### | 0.303### | – |
| | Average Response Time of Steering Wheel Angle (s) | Myo Armband | – | 0.038#### | 0.025#### |
| | | Steering Wheel | 0.038#### | – | 0.125#### |
| | | Manual Takeover | 0.025#### | 0.125#### | – |
| Pedestrian Not at Crosswalk | Minimum Average Distance to Pedestrian (m) | Myo Armband | – | 0.004#### | 0.002#### |
| | | Steering Wheel | 0.004#### | – | 0.278#### |
| | | Manual Takeover | 0.002#### | 0.278#### | – |
| | Maximum Average Steering Wheel Angle (º) | Myo Armband | – | 0.014## | 0.002## |
| | | Steering Wheel | 0.0137## | – | 0.290## |
| | | Manual Takeover | 0.002## | 0.290## | – |
| | Average Response Time of Steering Wheel Angle (s) | Myo Armband | – | 0.117#### | 0.250#### |
| | | Steering Wheel | 0.117#### | – | 0.028#### |
| | | Manual Takeover | 0.250#### | 0.028#### | – |

1. $* \ p < 0.01$
2. #$t$-test for equal variances
3. ## $t$-test for unequal variances
4. ###Wilcoxon signed-rank test

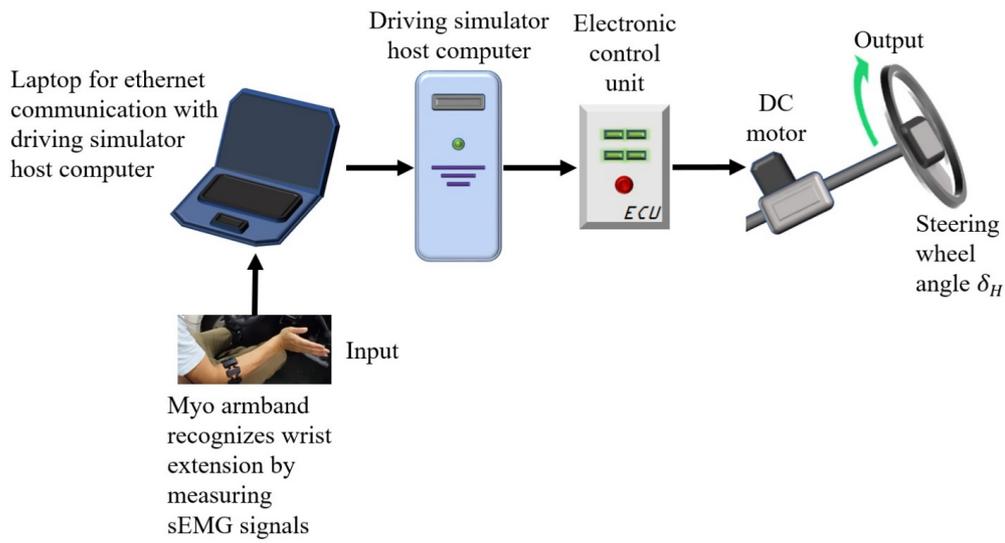

Figure 1. Overall control scheme of steering assistance interface.

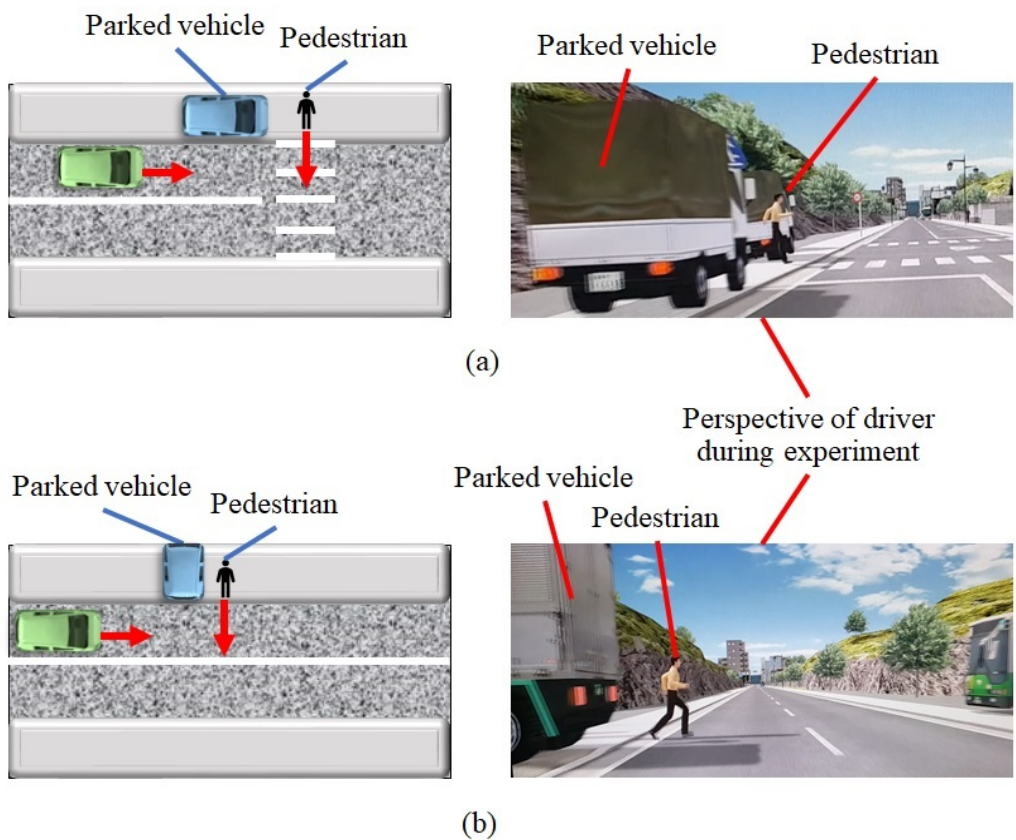

Figure 2. Driving scenarios with collision avoidance of pedestrian at crosswalk (a) and without crosswalk (b).

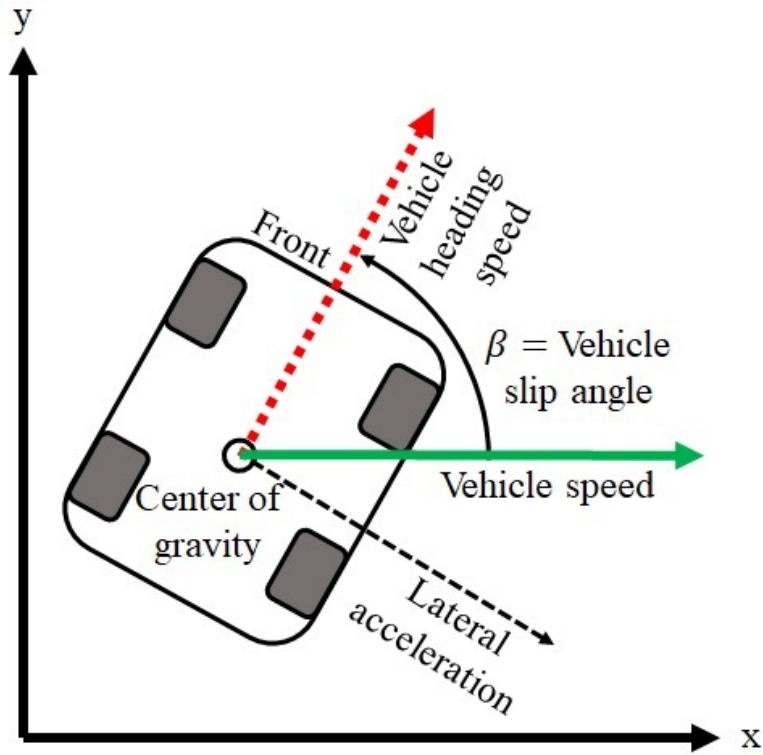

Figure 3. Vehicle slip angle, *β*, relative to center of gravity of simulated vehicle.

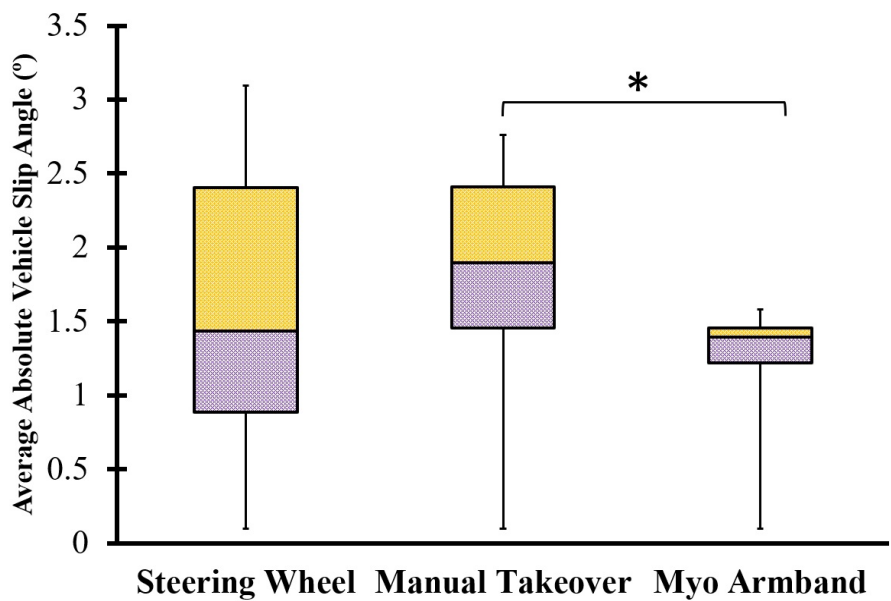

Figure 4. Comparison of steering interfaces across pedestrian collision avoidance scenarios. "*" indicates $p < 0.05$ based on Wilcoxon signed rank test.

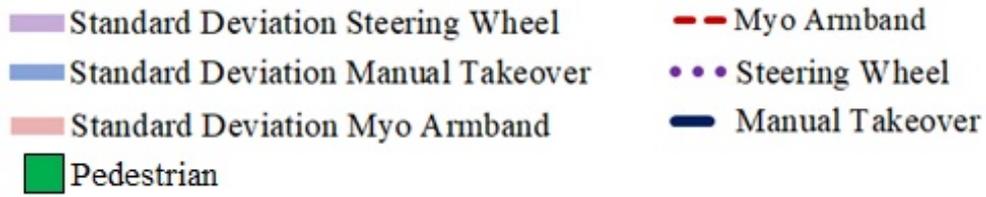

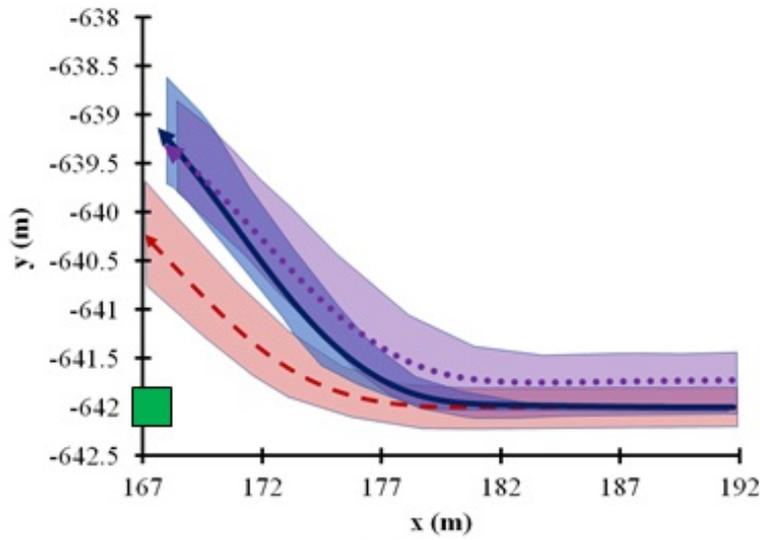

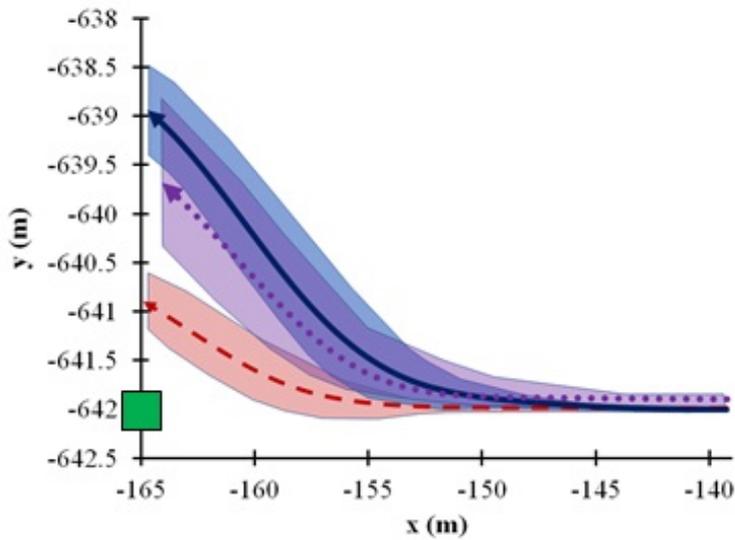

Figure 5. Comparison of average pedestrian collision avoidance trajectories over time in driving scenarios with a crosswalk (a) and without a crosswalk (b).

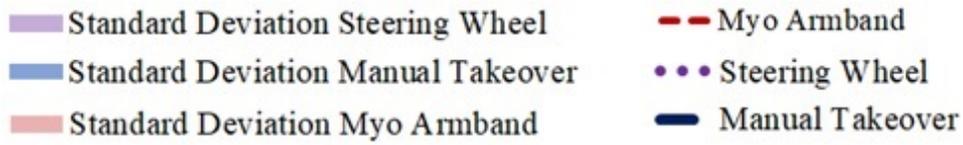

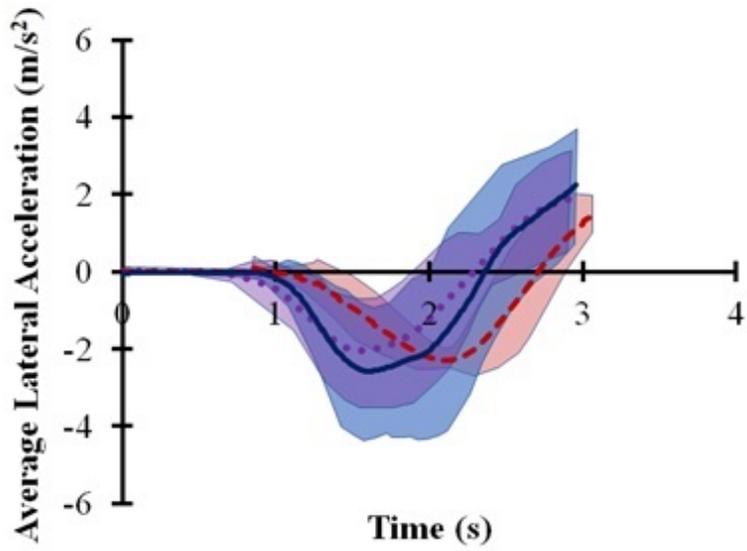

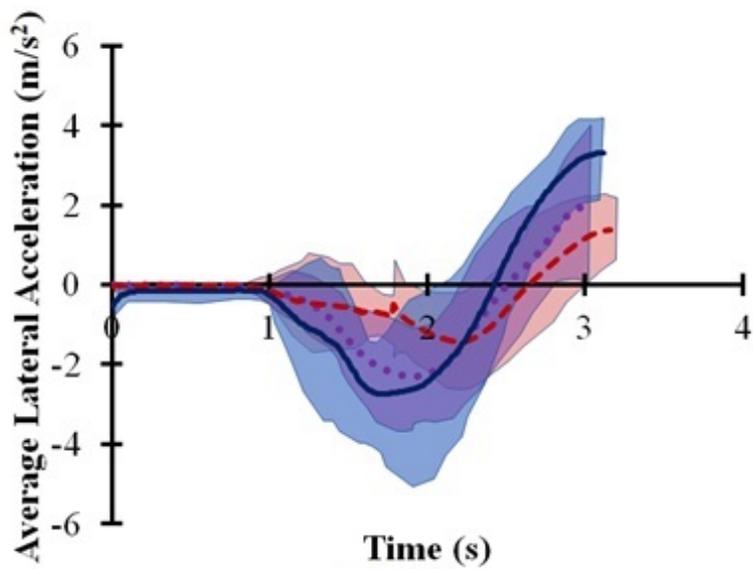

Figure 6. Comparison of steering interfaces with regard to average lateral acceleration over time in pedestrian collision avoidance scenarios with a crosswalk (a) and without a crosswalk (b).

# A list of figure captions

Figure 1. Overall control scheme of steering assistance interface.

Figure 2. Driving scenarios with collision avoidance of pedestrian at crosswalk (a) and without crosswalk (b).

Figure 3. Vehicle slip angle, $β$, relative to center of gravity of simulated vehicle.

Figure 4. Comparison of steering interfaces across pedestrian collision avoidance scenarios. "*" indicates $p < 0.05$ based on Wilcoxon signed rank test.

Figure 5. Comparison of average pedestrian collision avoidance trajectories over time in driving scenarios with a crosswalk (a) and without a crosswalk (b).

Figure 6. Comparison of steering interfaces with regard to average lateral acceleration over time in pedestrian collision avoidance scenarios with a crosswalk (a) and without a crosswalk (b).